\documentstyle[aps,prl,multicol,epsf]{revtex}

\begin{document}
\def\be{\begin{equation}}
\def\ee{\end{equation}}
\def\bc{\begin{center}}
\def\ec{\end{center}}
\def\bea{\begin{eqnarray}}
\def\eea{\end{eqnarray}}
\draft
\title{Packet Transport on Scale Free Networks}
\author{Bosiljka Tadi\'{c}$^1$ and G.J.Rodgers$^2$}

\address{$^1$Jo\u{z}ef Stefan Institute, P.O. Box 3000, 1001 Ljubljana,
Slovenia}
\address{$^2$Department of Mathematical Sciences, Brunel University,}
\address{Uxbridge, Middlesex UB8 3PH, U.K.}
\maketitle \thispagestyle{empty}
\begin{abstract}
We introduce a model of information packet transport
on networks in which the  packets are posted by a given rate and
move in parallel according to a local search algorithm.
By performing a number of simulations
we investigate the major kinetic properties of the transport as a function
of the network geometry, the packet input rate and the buffer size.
We find long-range correlations in the power spectra of arriving packet
density and the network's activity bursts. The packet transit time
distribution shows a power-law dependence with  average transit time
increasing  with network size. This implies dynamic queueing
on the network, in which many interacting queues are mutually driven by
temporally correlated packet streams.

{\it Keywords:} Internet, traffic, information packet
\end{abstract}
\pacs{PACS numbers: 02.50.cw, 05.40.-a, 89.75Hc.}
\begin{multicols}{2}

\section{Introduction}

The Internet has become a central feature of all our lives.
This has lead to interest in
information packet transport on  massive, heterogeneous, random,
network geometries.

Many empirical studies of packet transport on the Internet have been
carried out \cite{csabai,t1,chong,erramilli}. The principal
conclusion of these measurements is that the  aggregate
packet streams are fractal, obeying long-range correlated in time.
Of particular interest are
studies of packet density at a particular link (or on a node) and
ping time statistics \cite{t1,chong}, in which the round trip time,
or the time it
takes a packet to travel to a destination and back to its source,
are measured.
Analysis of the power spectrum of the packet density and
round trip times allows one to distinguish two regimes with {\it free flow}
and {\it jammed} traffic, respectively, depending on the traffic intensity.
In \cite{erramilli} it was shown that the power-law
behaviour of the distribution of packet inter-arrival times
has a significant effect on packet queuing performance, and
consequently on the overall packet transport.

These dynamical transport and queuing  processes are taking
place on the Internet, the network made up of
routers and computers as vertices and cables as edges.
Recent studies of the geometry of the Internet,
\cite{fal,gov,Vazquez}, indicate that the
degree distribution has a power-law, scale free behaviour
\cite{ba,DMS,BT,KRR}.

Another topological property, the betweenness, or the total number of
shortest paths going through each node \cite{load,newman},
was also found to have a power-law
distribution with an exponent $\approx 2.2$ on a scale free graph.
In terms of transport processes,
this corresponds to the distribution of the number of packets
transferred through  a node (in a long-time limit),
in a system in which the packet transport is
dominated by non-interacting packets that always
take the shortest route between source and destination.

In reality, the information packet transport is much more complex for
the following reasons: (1)  no global navigation is technically available;
(2) many packets are being transported simultaneously, hindering each others
motion, depending on the search algorithm and network structure.
In particular, the kinetics of many interacting packets leads to a
qualitatively new feature, which is manifested in the {\it queueing}
\cite{book} of
packets at individual nodes. In the network we have the problem of many
interacting queues.
Thus a close relationship seems to exist between the kinetic properties of
the packet transport, such as the distribution of round trip times, or
the number of packets on a node, and the network on
which the kinetics takes place. The interplay between network structure
and packet kinetics becomes particularly important when
one imagines the transport in a network with buffers, which restrict the
length of packet queue at each node, or with reduced ability of nodes
to handle the packets.
In recent studies of traffic on idealized geometries---on a linear chain
\cite{micro} and on Cayley trees \cite{Arenas,Guimera} and one- and
two-dimensional lattices \cite{Guimera} a sharp transition from a free to
congested traffic was found.
The smallest buffer along the chain causes jamming in the linear model
\cite{micro}, while transport through the node at the top of hierarchy is
crucial for the onset of congestion on the hierarchical lattice
\cite{Arenas,Guimera}.
In a network with scale-free structure of links the multiplicity of
potential paths among pairs of nodes may alter the role of buffers in
the crossover to jammed traffic. It is expected that
  the local geometry of the hub nodes, and their buffer sizes,
is of critical importance, while the buffer sizes of the
majority of other nodes  are largely irrelevant.
The relative importance of nodes,
of course, depends on the search algorithm being used.
For practical reasons,
search algorithms can only be applied to small sections of the network.
An attempt was reported recently \cite{Albert} to find optimal topology of
the network for a given local algorithm in the presence of congestion.

In this paper we introduce a simple model of packet transport on
networks in an attempt to understand which of the elements present in a
real system are necessary for the system to display the observed empirical
results. In particular, we study the effect of the
geometry of the network on the major
properties of the packet transport. We develop a model for simultaneous
transport of  packets
and implement this model on scale free  and randomly grown networks.
Packets are  posted by a given input rate $R$ and moved according to
a local algorithm which uses up to next-neighbour search
and the local geometry. In this way the packet queue at each individual
node is dynamically formed by packets  moving from neighbouring
nodes. We measure a number of kinetic quantities that allow us to
characterize the nature of the packet transport process on the two networks.

The paper is organized as follows: In Section II we introduce the model by
first presenting the algorithm of graph growth and then the search algorithm
for packet kinetics on the graphs. In Section III we define precisely the
quantities that we determine in the simulations and present the results
for the packet time series and power spectra of these quantities.
In Section IV we study the distribution of transit times of packets,
the queue lengths and the networks' output rate.
In Section V we present a short summary of the results and
discuss the main conclusions.

\section{The Model}

 Our model is developed in two separate stages.
Firstly the network is grown and then
we simulate the motion of packets on the network.

{\bf Network Structure}

We consider two different tree networks, one of which is scale-free and
the second is a grown random network.

The {\bf scale free} network (SFN) structure \cite{ba} is grown as follows.
At each time step one new node is added to the network and is linked to
a node of degree $k$ with a rate \cite{BT}
\begin{equation}
p(k,t)= \frac{k+\alpha}{D(t)+\alpha N(t)}.
\end{equation}
Here $D(t)$ is the total degree of the network at time $t$ and
$N(t)$ is the total number of nodes in the network, and  $\alpha >0$
is a tunable parameter. For networks grown in this way, the
degree distribution is power-law \cite{ba,BT,DMS,KRR}. More precisely,
the number
of nodes with degree $k$, $P(k)$, behaves as $P(k) \sim k^{-\tau}$ with
$\tau =2+\alpha$.
Choosing $\alpha=0.2$ means that this network has
the same in-degree distribution \cite{fal} as that observed on the Internet.

The {\bf random network} (RGN) is grown by adding one new node at each
time step
and connecting it to a node of degree $k$ with rate $1/N(t)$, independent of
$k$. In this case the degree distribution $P(k)$ behaves as $P(k)\sim 2^{-k}$.

{\bf Sending Packets}

At each time step, with probability $R$, a new information
packet is initiated.
This is done by randomly selecting a source node and a
destination node for that packet.

{\bf Packet Transport}

At each time step, each node is investigated in turn, and if it has
a packet on it then the top packet on the node attempts to move.
This is done by searching the nearest and next nearest neighbours
of the node for the destination node of that packet. If found, the packet is
moved to or towards the its destination node.
If the destination node isn't
found, the packet moves to a randomly selected neighbour.
Each node has a buffer size of integer $B$, the maximum
number of packets that can be on a node at any one time.
If a packet attempts to move to a full node, it is unable to stay there
and moves back to the node it came from.
We assume that every node has the same buffer size.
Similarly, in order to fully investigate the
effect of the geometry of the graph, we assume that  all the links have
the same capacity. When the packet reaches its destination node, it is
removed from the network.

In terms of queuing theory \cite{book},
our nodes can be thought of as single server
queues, with buffer size $B$, and a deterministic service time distribution.
The service discipline is last in first out (LIFO). For single server
queues with this discipline some
approximations have been developed \cite{abate} for the waiting time
distribution.
Finally the inter-arrival time distribution is not put in by hand, but
arises naturally  as a function of
the network geometry and the kinetics of the packets at neighbouring  nodes
on the network.
The waiting time distribution has two different contributions: First,
a node receives packets with a Poisson distribution
as a result of packets being initiated on that node;  and, second,
the node receives packets from neighbouring nodes
with a time series that turns out to have a long range, power-law
temporal correlation.

Thus we have packet transport rules which yield paths for the packet
which are close to, but longer than, the minimum path between source and
target node. We also have networks which are trees containing no loops.
We do not believe that the loops present in real networks \cite{fal,BT,KRR}
play an important role in the observed behavior
\cite{csabai,t1,chong,erramilli}.

\section{Traffic Time Series and Power Spectra}

We have performed a number of simulations of the kinetics of the packet
transport on both the scale free network and the random grown network of size
$N=10^3$ nodes.
We considered a number of different quantities that characterise the
kinetics and allow comparison to be made
with real data from the Internet \cite{csabai,t1}. These were

(a)	Activity, $a(t)$,
	the total number of nodes with a packet on at time $t$.
This gives a measure of the fraction of the network that is busy.

(b)	Total Load, $n(t)$, the total number of packets
	in the network at time $t$.

(c)	Load (or queue length), $q(t)$, the number of packets
on a node at time $t$.

(d)	Load on the active nodes at time $t$, $l_a(t)= \sum _{active}q(t)$.
This quantifies the load carried by the active nodes at a particular time step.

(e)	Density, $\rho(t)$,
	the total number of packets arriving at the hub node at time $t$.

(f)	Transit time, $T_{tr}$, the time taken by a packet to reach
its destination.

The characteristics of these quantities change as a function of the geometry
of the network, the buffer size $B$ and the input rate $R$. In order to
examine these changes, we calculated their power-spectra. In general,
the power spectrum of a time series $X(t)$ is defined by
\begin{equation}
S_X(f)=|\sum_{t=1}^{\infty}X(t)e^{ift}|^2.
\end{equation}
For a time series with no
temporal correlation, the plot of $S_X(f)$ against
$f$ is independent of $f$.
If the time series $X(t)$ has an auto-correlation function
$r_X(k)\sim k^{-\phi}$ then $S_X(f)\sim f^{-\phi}$.

The output rate, or average number of packets leaving the
system per time step, is, along with the transit time $T_{tr}$, a
measure of the efficiency of the system. It is defined by
\begin{equation}
\mu = R-\frac{n(t)}{t}.
\end{equation}
In general the value of the ratio $\mu /R$ allows one to determine
\cite{Whitt} whether the queues in the system are getting longer
with time. As it will be shown below, in our model, the queues
are growing with time for all values of the parameters.

A comparison of the packet traffic on the scale-free network and the
randomly grown network in our simulations is
demonstrated by time series for load carried by active nodes
and by number of nodes that are active at given time  shown in Fig.\ 1.
With identical driving conditions,
the random network tends to distribute the
activity over many nodes simultaneously, whereas in the scale-free
network the activity is localised to the hub, which carries most of
the packets, and a few other high degree nodes in the centre of the network.

In Figure 2 we compare the power spectra of the time series for activity,
load at active nodes, density at the hub, and number of packets in the network
for the two network geometries  operating in the identical driving conditions
of low input rate $R=0.01$ and buffer size $B=100$  for network size
$N=1000$ nodes.
For these driving conditions the long-range correlations in the time series
are present in certain range of frequencies in both networks. The slopes are
generally closer to -1 in the case of SFN  ($1/f$-noise), whereas in the RGN
we measure the slope close to -2, except for the density where it is close
to -0.5, and a whiter spectrum (c.f. Fig.\ 2).
These features, however, change with increased input rate and/or decreased
buffer size, suggesting that the character of the transport
depends  on the relative ratio of these parameters.

In particular, when the input rate is increased by a factor of four and the
buffer size  is unchanged ($B=100$, $R=0.04$) the  power spectrum of the
density at the hub in SFN behaves as $1/f^2$ (top line in Fig.\ 3).
In the case of RGN the slope becomes approximately -1 and the range
of frequencies where the correlation occurs is reduced.
Changes in the spectrum of density indicates that the character of packet
transport at the hub changed from $1/f$ noise in the free-flow regime,
to simple diffusive transport.

In Figure 3 we also show
the power spectrum of the density measured at a node away from the hub
in the free-flow regime for both network geometries. It is interesting
to note that, in the SFN, in contrast to the density at hub, the spectrum
inside the network for low traffic intensity is a white noise
(bottom curve in Fig.\ 3). In the
RGN, however, the correlations similar to the ones at the hub are measured,
once again supporting the conclusion that the RGN tends to distribute
the activity over  the entire graph.

\section{ Packet Transit Times and Queueing Properties}

To further investigate how the network performs under dense packet traffic
we study transit times for individual packets. Within our algorithm we
mark the first $10^3$ packets with an additional time label and monitor the
time that they spend on the network
from posting until arrival at their respective destinations. Packets are
still posted probabilistically with rate $R$. In the inset to Fig.\ 4 an
example of the time series of transit times against posting time
of the packets is shown. It is noticeable that
the duration of the journey for each packet
is different, both because the difference in distances that they make on
the graph and because of the time that they spend waiting in queues at
particular nodes.
The distribution of transit times is given in main part of Fig.\ 4 both
for SFN and RGN for low packet density ($R=0.01$, $B=100$).

For the range of values of the transit times $T_{tr}$ we find
power-law distributions, but with different slopes, indicating the
difference in the efficiency of the two network structures.
Namely, $\tau_T=1.25 \pm 0.02$, for SFN, and $\tau_T=0.53\pm 0.02$, for RGN.
In addition, the transit times larger than approximately $10^3$ time steps
for SFN, and $10^4$ in the case of RGN, contribute to  tail of the
distribution.
Among these are packets that become buried deep in  queues at certain
nodes in the network. Thus, it is the dynamics of queueing, in addition to
the network's structure, that determines the network's transport efficiency.
Next we study some properties of queues in both networks.

In Figure 5 we show the probability distributions of the queue lengths
(loads) measured at individual nodes throughout the network {\it after}
each time step is
completed. Apart from the cut-off which is determined by the maximum
buffer size (here $10^3$),  there is a larger
probability to find a queue of given length $q$ in the RGN compared to the SFN,
in agreement with generally lower efficiency of the randomly grown network.
The relative appearance
of large queue lengths in the RGN decays with the power-law exponent
 $\tau _q=1.4 \pm 0.02$, and with $\tau_q = 0.48 \pm 0.02$ in the SFN.
In both cases there is a part of the network in which nodes carry a small
load, with a similar power-law decay.

Compared to standard single-queue theories \cite{book}, here we have
many interacting queues in which packets hop from one queue to the other along
the edges of the graph, as they are directed by the search algorithm.
Thus the input rate of packets at each node is different and related to the
local connectivity of that node and number of paths that the actual search
algorithm selects to pass through that node at a single time step.
(As already mentioned in the introduction, we keep the output rate of one
packet per time step equal at each node, in order to single out the effects
of networks structure more clearly.)
Therefore, in the SFN, the kinetics of the queues at the nodes with high
 connectivity are  of primary importance, whereas queues at the
majority of other nodes in the network rarely exceed a few packets.
This is due to the topology of SFN, in which cluster of nodes linked to
the hub increases faster than any other subgraph,
 and also due to the search algorithm we are using, which is suitable for that
topology. In the RGN the same search algorithm is much
less effective, because the dominant cluster grows only logarithmically
in time making the edges  more evenly distributed throughout the network.

Thus, the input rate at the hub varies in time and it is precisely given by
the density $\rho (t)$ that we determined in Section\ III. In the SFN
$\rho (t)$ was found to exhibit long-range correlations of $1/f$ type
for low intensity of traffic (cf. Fig.\ 2), that changes to a short-range
$1/f^2$ correlations when traffic intensity increases. In the RGN within
the same conditions, for comparison, the spectrum is much closer to white
noise. Next we investigate how the queueing dynamics at  the hub affects
performance of the network.

We determine the network's output rate $\mu$, defined as the number of
 packets arriving to their destination per time step. Therefore, the
difference of the input and output rate $R-\mu = n(t)/t$  determines the
workload \cite{Whitt} on the level of the entire network.
In main part of Fig.\ 6 we show the workload
in SFN for two input rates $R$ and fixed buffer size $B=100$.
For large input rate
(top curve) $R-\mu$ saturates for large times $t$ at a finite average value,
indicating that total load in the network  $n(t) =(R-\mu)t$ increases
{\it linearly} in time. In this case the jamming that first occurs at
the hub seems to spread gradually throughout network.
For comparison, in the case of low input rate (bottom curve) the
workload $R-\mu$, after a sharp initial increase, decays
approximately as $\sim t^{-1/2}$, indicating that total load $n(t)$
increases  {\it sub-linearly} in this flow regime.
Taking the average asymptotic value of $\mu$, in the inset to Fig.\ 6 we
show the ratio of the network's output and input rates $\mu /R$
for the SFN and the RGN, which, once again shows the systematically better
performance of the scale-free structure for a range  of input rates.

\section{Conclusions and Discussion}

In this paper we have introduced a new model of packet transport on
networks and investigated its major properties on
both randomly grown and scale free networks.
The model incorporates in a natural way three basic elements---network
topology, search algorithm, and queueing dynamics---which determine
overall efficiency of the transport. By fixing the input rate and using the
 near and next near neighbour search algorithm,
which is fairly efficient on structured graphs, in this work  we
focused on the effects of networks' topology on the queueing dynamics.
For this purpose we also keep link capacities at a minimal one packet per
time step.

We found long-range temporal  correlations for many of the important
quantities within the system, in particular for the density of arriving
packet streams, with
non-universal exponents that were dependent on the traffic intensity and
the network on which the transport was taking place.
The power-law dependences of the transit time distribution with
the exponent $\tau_T < 2$ implies that the average time that a packet
spends on the network increases with network size, causing queueing
processes on the network. We find the power-law behavior
of the queue length distribution reflects the networks' structural
efficiency in the transport.
The character of the queueing dynamics in our model is dominated by
queues  at highly connected nodes, which are driven by fractal packet
streams from neighbouring nodes. Measured by the single-queue criteria,
the queues in our model are predetermined to increase with time.
However, measured on the level of entire network, we find that the networks'
workload may increase sub-linearly for low traffic intensity, and linearly
when traffic approaches the congested regime.
At the same time the observed long-range correlations
and power-law dependences of the distributions are different in two
traffic regimes, a transition seems to take place along a line $R_c(B)$.
A detailed analysis of this transition was not included in this work.

Finally we found that the scale free network was much more
efficient, in terms of network output to
input ratio, than the randomly grown network. Allied to
this was the observation that transport on the scale free network was
very dependent on the queuing at a few highly connected nodes, and many of
the low-connectivity nodes had very low activity. In contrast,
the randomly grown network distributed the activity much more
evenly over the graph.

This work is an initial study, in future we intend to investigate a
system of this type with (i) different search algorithms and (ii) different
queuing disciplines, with a view to
investigating their effect on packet transport.

\section{Acknowledgements}
We would like to thank NATO for partial financial support through PEST grant
PST.CLG.978404. Work of B.T. was  supported by the Ministry
of Education, Science and Sports of the Republic of Slovenia.


\end{multicols}
\newpage \twocolumn
\narrowtext
\begin{figure}
\epsfxsize=80mm\epsffile[46 72 514 486]{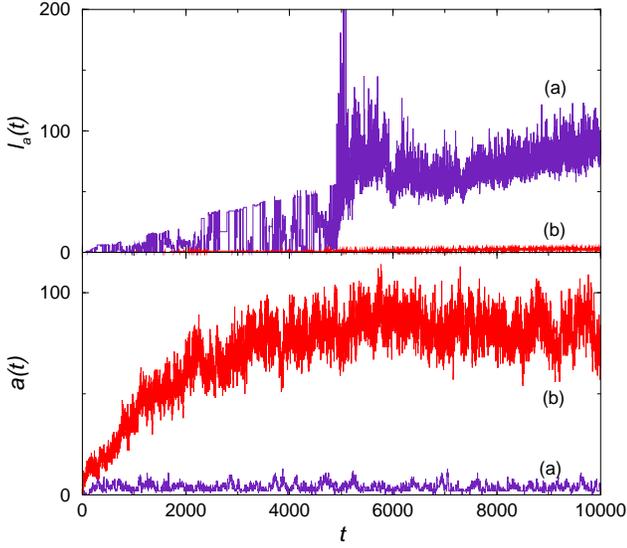}
\vskip 0.8true cm
\caption{\label{fig1} Average load at active nodes $\ell _a (t)$
vs. time $t$ (top panel) and number of active nodes in the network $a(t)$ at
time $t$ (lower panel) in jamming regime $R=0.04$, $B=100$ for scale-free
graph (curve a) and for randomly-grown graph (curve b) with the same
 driving conditions.
 }
\end{figure}

\begin{figure}
\epsfxsize=80mm\epsffile[43 71 473 615]{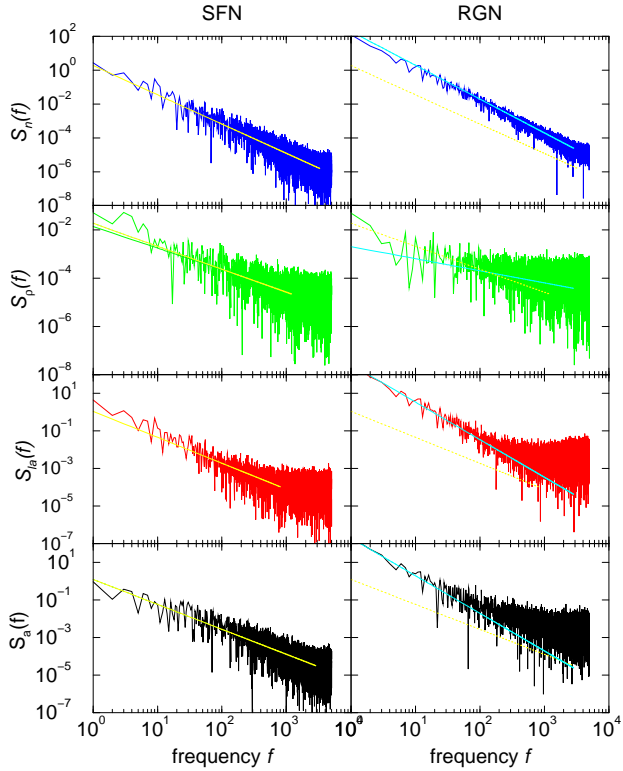}
\caption{\label{fig2} (bottom to top) Power spectrum of activity,
load on active nodes, density at hub, and total load for scale-free
graph (left column) and in randomly grown graph (right column) for
driving conditions corresponding to the free flow regime ($R=0.01$, $B=100$).
Power-law fits are shown by solid lines. Copies of the fits for SFN are
also shown in the corresponding panels for RGN as dotted lines to
display the differences in slopes and in the correlation ranges. }
\end{figure}

\begin{figure}
\epsfxsize=80mm\epsffile[63 96 505 546]{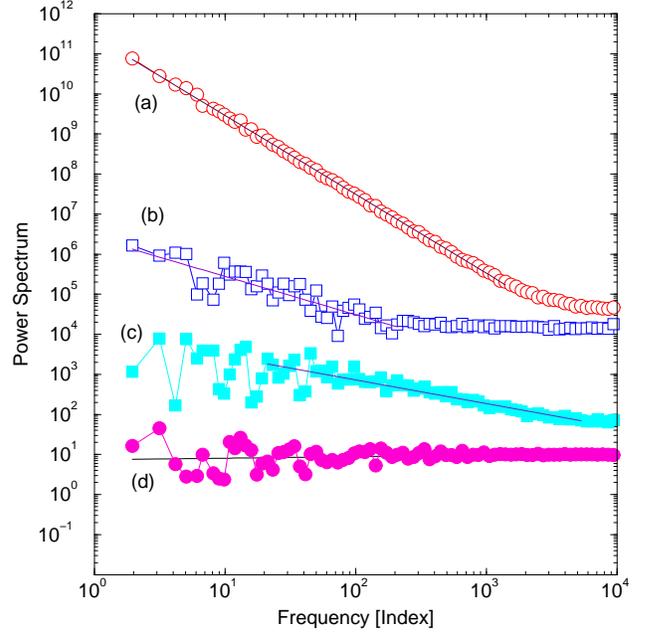}
\vskip 0.8true cm
\caption{\label{fig3}Power spectrum of the density measured at the hub for
$R=0.04$ and $B=100$ in (a) SFN and (b) RGN, and at far node in
free-flow regime ($B=1000$) in  (c)  RGN  and (d) SFN. Curve (c)
has been shifted shifted to display it more clearly and data
has been log-binned. Fitted slopes are (top to bottom) -2, -1, -0.59,
and 0.03.}
\end{figure}

\begin{figure}
\epsfxsize=80mm\epsffile[70 70 508 488]{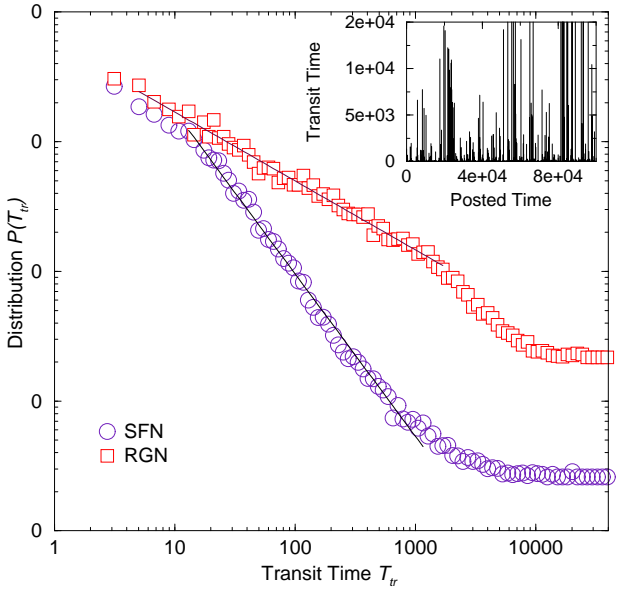}
\vskip 0.8true cm
\caption{\label{fig4}Probability distribution of transit times of
marked packets in the conditions of free flow ($R=0.01$,$B=100$).
 for RGN and SFN. Log-binning ratio 1.1 . Inset: A set of transit
times of 1000 marked packets against posting time, for the same
conditions as in the main figure.
  }
\end{figure}

\begin{figure}
\epsfxsize=80mm\epsffile[44 71 508 488]{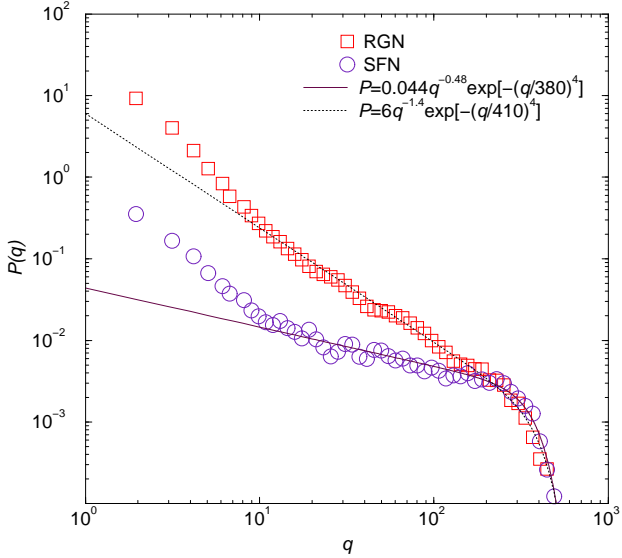}
\vskip 0.8true cm
\caption{\label{fig5}Probability distribution of the queue lengths
 counted after each time step was completed for RGN and SFN and the
respective fit lines.
Driving rate $R=0.12$ and buffer size  $B=1000$. Log-binning ratio 1.1 .
 }
\end{figure}

\begin{figure}
\epsfxsize=80mm\epsffile[43 71 507 488]{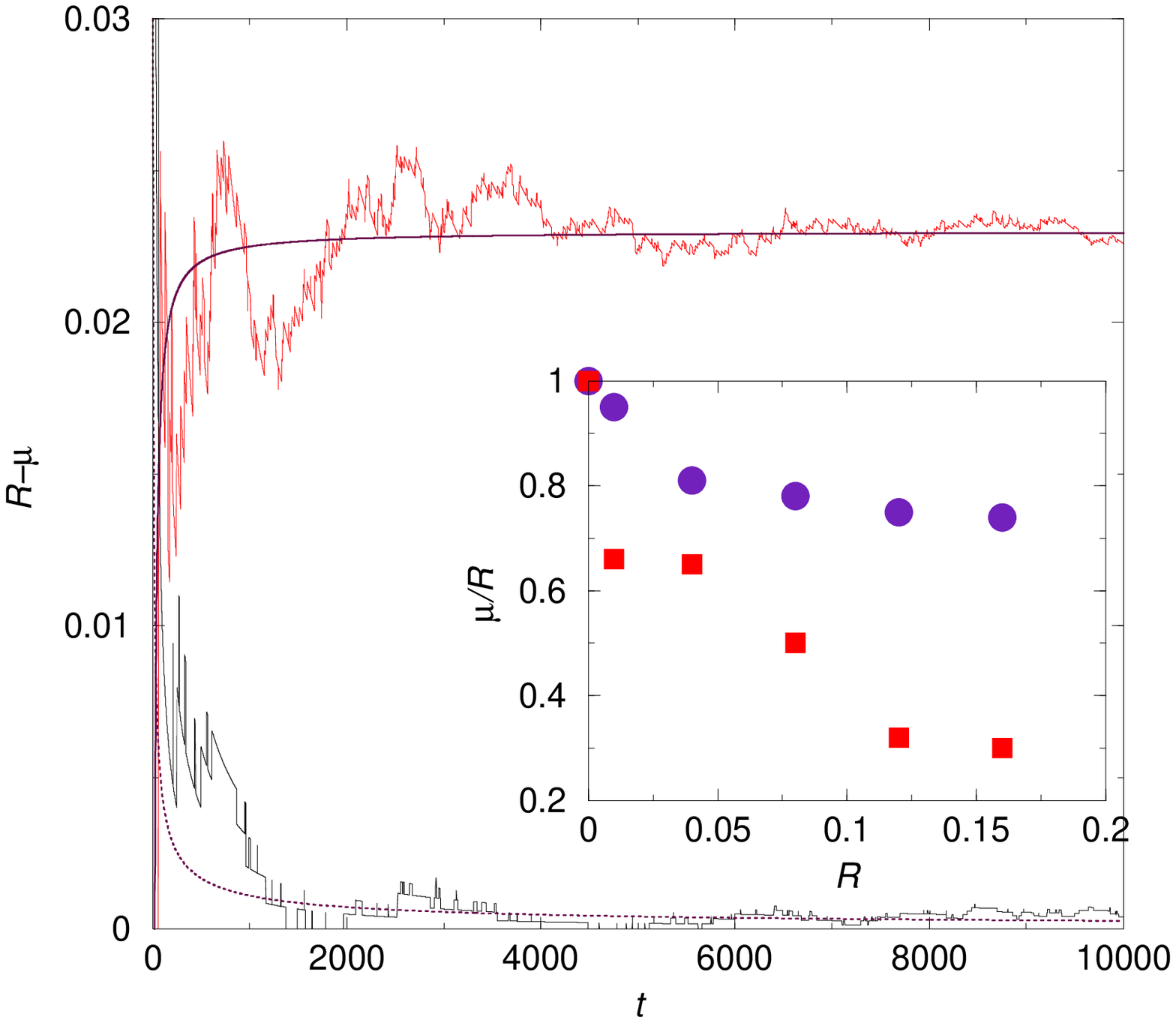}
\vskip 0.8true cm
\caption{\label{fig6}Time dependence of the workload $n(t)/t$ in SFN for two
input rates $R=0.08$ (top) and $R=0.01$ (bottom) and buffer size $B=100$.
Fits: $R-\mu = 0.023 - 0.5t^{-1}$ (solid line) and $R-\mu = 0.07t^{-0.5}$
(dotted line).
Inset: Average output rate of the network $\mu /R$ normalized to input
rate $R$, vs $R$ for SFN (bullets) and RGN (squares).}
\end{figure}


\begin{thebibliography}{99}
\bibitem{csabai}I. Csabai, J. Phys. A {\bf 27} 417 (1994).

\bibitem{t1} M. Takayasu, H. Takayasu and T. Sato, Physica A {\bf 233} 824
(1996).
\bibitem{chong}K. B. Chong and Y. Choo, physics/0206012.
\bibitem{erramilli} A. Erramilli, O. Narayan and W. Willinger,
IEEE/ACM Trans. Networking {\bf 4} 209 (1996).


\bibitem{fal} M. Faloutsos, P. Faloutsos and C. Faloutsos, Proc. ACM SIGCOMM,
Comput. Commun. Rev. {\bf 29} 251 (1999).
\bibitem{gov} R. Govindan and H. Tangmunarunkit, Proc. IEEE Infocom 2000,
Tel Aviv, Israel (2000).
\bibitem{Vazquez} A. Vazquez, R. Pastor-Satorras, and A. Vespignani,
cond-mat/0112400.

\bibitem{ba} R. Albert and A.-L. Barabasi, Rev. Mod. Phys. {\bf 74} 47 (2002).
\bibitem{DMS}S.N. Dorogovtsev,  J.F.F. Mendes, and   A.N. Samukhin,
Phys. Rev. Lett. {\bf 85}, 4633 (2000).
\bibitem{BT} B. Tadi\'{c}, Physica A {\bf 293} 273 (2001).
\bibitem{KRR} P.L. Krapivsky, G. J. Rodgers, and S. Redner,
Phys. Rev. Lett. {\bf 86}, 5401 (2001).

\bibitem{load} K.-I. Goh, B. Kahng and D. Kim, Phys. Rev. Lett. {\bf 87}
278701 (2001).
\bibitem{newman}M. E. J. Newman, Phys. Rev. E {\bf 64} 016131 (2001).

\bibitem{book} G. Gross and C.M. Harris, {\it Fundamentals of Queuing Theory},
(Wiley, New York 1998).


\bibitem{micro} T. Huisinga, R. Barlovic, W. Knospe, A. Schadschneider and M.
Schreckenberg, Physica A {\bf 294} 249 (2001).

\bibitem{Arenas}A. Arenas, A. Diaz-Guilera, and R. Guimera, Phys. Rev. Lett.
{\bf 86}, 3196 (2001).
\bibitem{Guimera}R. Guimera, A. Arenas,  A. Diaz-Guilera, and F. Giralt,
cond-mat/0206077.

\bibitem{Albert}R. Guimera, A. Arenas, A. Diaz-Guilera, F. Vega-Redondo, and
A. Carbales, cond-mat/0206410.

\bibitem{abate} J. Abate and W. Whitt, Opns. Res. Letters {\bf 20} 199 (1997).
\bibitem{Whitt} W. Whitt, {\it Stochastic-Process Limits}, Ch. V ,
(Springer, New York 2002).

\end{thebibliography}
\end{document}